\documentclass[twocolumn,superscriptaddress,nopreprintnumbers,amsmath,pra]{revtex4}
\usepackage{graphicx}
\usepackage{amssymb}

\begin{document}
\title{Comparison of discharge lamp and laser pumped cesium magnetometers}
\author{S. Groeger}
\affiliation{Physics Department, Universit\'e de Fribourg, Chemin de
Mus\'ee 3, 1700 Fribourg, Switzerland} \affiliation{Paul Scherrer
Institute, 5232 Villigen PSI, Switzerland}
\author{A.~S. Pazgalev}
\affiliation{Physics Department, Universit\'e de Fribourg, Chemin de
Mus\'ee 3, 1700 Fribourg, Switzerland} \affiliation{Ioffe Physical
Technical Institute, Russ. Acad. Sc., St. Petersburg, 194021,
Russia}
\author{A. Weis}
\affiliation{Physics Department, Universit\'e de Fribourg, Chemin de
Mus\'ee 3, 1700 Fribourg, Switzerland}

\date{\today}

\begin{abstract}
We have performed a comparison of laser (LsOPM) and lamp (LpOPM)
pumped cesium vapor magnetometers. Although the LsOPM operated 50\%
above its shot-noise limit we found an intrinsic sensitivity of
15\,fT/$\sqrt{\mbox{Hz}}$ and 25\,fT/$\sqrt{\mbox{Hz}}$ for the
LsOPM and LpOPM respectively. Two modes of operation, viz.,~the
phase-stabilized and the self-oscillating mode were investigated and
found to yield a similar performance. We have compared the
performance of the LsOPM and the LpOPM directly by simultaneous
measurements of field fluctuations of a $2\,\mu\mbox{T}$ magnetic
field inside a multilayer magnetic shield and have used one of the
magnetometers for an active field stabilization. In the stabilized
mode we found a gradient instability of 25\,fT within an integration
time of 100\,s, which represents an upper limit of the long-term
stability of the magnetometers.

\end{abstract}

\maketitle

\section{Introduction}
\label{sec:intro}

The precise measurement and control of magnetic fields and field
fluctuations is of crucial importance in many fundamental physics
experiments. The suppression of systematic uncertainties in
experiments searching for permanent electric dipole moments (EDMs)
in atoms and neutrons is one prominent example. New generations of
EDM experiments with ultracold neutrons (UCNs) aim at a higher
statistical precision by the use of higher UCN flux and larger
storage volumes. In such experiments magnetic field drifts and
fluctuations can produce false EDM signatures thereby putting
stronger constraints on the magnetic field control. Although
magnetometers based on superconducting quantum interference devices
(SQUIDs) are the most sensitive magnetometers available to date they
are of limited interest for monitoring magnetic fields in large
volumes. Moreover, SQUIDs do not measure absolute field values.

Two distinct magnetometric techniques were used in past EDM
experiments. In the ILL experiment\cite{ILLEDM}, which has produced
the presently lowest upper bound on the neutron EDM, a vapor of
$^{199}$Hg atoms filled into the ultracold neutron storage chamber
(20 liter) served as ``cohabitating'' magnetometer. The PNPI
experiment \cite{Alt96}, on the other side, used a set of two
self-oscillating cesium vapor magnetometers placed above and below
the storage chamber for monitoring the field in the chamber. Both
techniques have pros and contras. Co-magnetometers yield only a
volume-averaged field value, which yields no information on field
gradients and their fluctuations. External magnetometers, on the
other hand, do not measure the field in the volume of interest
directly, but allow - if used in sufficient number - to access field
distributions, thereby permitting the active control of specific
multipole moments of the field. Borisov et al.~ have proposed a
large volume (external) magnetometer based on nuclear spin
precession in $^3$He \cite{Heil_He3Mag}. That device uses a double
pulse Ramsey resonance technique, which besides its lack of spatial
resolution also suffers from a lack of temporal resolution.

The PNPI experiment used two conventional state-of-the-art discharge
lamp pumped self-oscillating cesium vapor magnetometers
(OPM)\cite{Alt96}. Such types of magnetometers - developed since the
1950's - have a shot-noise limited performance and large bandwidths.
The high spatial and temporal resolution of optically pumped alkali
magnetometers thus make such devices interesting alternatives for
the continuous monitoring of fields, gradients, and fluctuations
thereof. The use of alkali OPMs for the field control in larger
volumes calls for a substantially larger number of sensor heads,
which suffers from the fact that a single discharge lamp can only
drive a limited number of sensors. The steady development in the
past decades of narrow-band semiconductor diode lasers makes such
light sources attractive alternatives to discharge lamps. Owing to
the high spectral density of its radiation a single diode laser of
moderate power (a few mW) can be used to drive dozens of
magnetometer heads.

Having a multichannel external magnetometer approach for a planned
neutron EDM experiment in mind we have performed a comparative study
of state-of-the-art discharge lamp pumped magnetometers (LpOPM) and
laser pumped magnetometers (LsOPM) using similar room temperature
sensor cells (7\,cm and 6\,cm diameter respectively) and identical
electronics. We discuss the principle of operation and details of
their practical realization. The devices were operated in two
distinct modes, viz., the self-oscillating mode and the
phase-stabilized mode. Details of the development and performance of
the LsOPM have been published earlier \cite{FRAPLsOPM}. We have
determined the intrinsic sensitivities of the magnetometers and
present measurements of the fluctuations of a $2\,\mu\mbox{T}$ field
recorded simultaneously by the LsOPM and the LpOPM in a multi-layer
magnetic shield. The LpOPM reached its ultimate shot-noise limited
performance while the LsOPM showed a superior intrinsic sensitivity,
although its performance still lies 50\% above its fundamental
shot-noise limit.

\section{Optically pumped magnetometers}
\subsection{General principle}

An OPM measures the Larmor precession frequency $\omega_L$ of a
vapor sample of spin polarized atoms in an external magnetic field
$B_0$. In small magnetic fields
\begin{equation}\label{eq:Larmorfrequency}
    \nu_L=\frac{\omega_L}{2 \pi}=\frac{\gamma_\mathrm{A}}{2\pi} B_0\,,
\end{equation}
and the field measurement reduces to a frequency measurement. In
Eq.~\ref{eq:Larmorfrequency} the subscript of $\gamma_\mathrm{A}$
refers to the total angular momentum of the precessing atomic state.
Although OPMs based on nuclear spin polarization were demonstrated
in the past
\cite{ILLEDM,Heil_He3Mag,MoreauGilles_He3LsMag,Gilles_He4LsMag_2001}
we restrict the present discussion to alkali vapors in which the
precessing levels are one or both of the hyperfine ground states
with total angular momentum $F=I\pm 1/2$, where $I$ is the nuclear
spin. Magnetic resonance is used to measure the precession frequency
by inducing resonant spin flips by a weak magnetic field $B_1$
oriented at right angles with respect to $B_0$ and oscillating at
the frequency $\omega_\mathrm{rf}$. Although for the magnetometer
discussed here $\omega_\mathrm{rf}$ lies in the audio range of
frequencies the index ``rf'' (radio-frequency) is used to comply
with common notation.

Optical pumping with a resonant circularly polarized light beam
creates spin polarization in the medium (room temperature alkali
atom vapor contained in a glass cell) and hence an associated net
bulk magnetization. It has been realized for many years that the
pumping process is most efficient for $D_1$ resonance light driving
the transition $|nS_{1/2}\rangle\rightarrow |nP_{1/2} \rangle$,
although magnetometers can also be realized using $D_2$
($|nS_{1/2}\rangle\rightarrow |nP_{3/2} \rangle$) light. In general
the optically pumped medium becomes transparent with respect to the
pumping light, except for the spectrally resolved closed
$|nS_{1/2}\,F\rangle\rightarrow |nP_{3/2},F+1 \rangle$ transition
\cite{Kazantsev84}, in which case the absorption of the pumped
medium increases. The fact that the optical properties of the medium
depend on its spin polarization is used to detect the magnetic
resonance transition by monitoring either the power or the
polarization of the transmitted or scattered light beam. The
technique is known as optically detected magnetic resonance (ODMR).

In the present study we have used a particular realization of the
ODMR technique, the so-called $M_x$-method, in which $B_0$ is
oriented at 45$^\circ$ with respect to the direction of propagation
($\widehat{k}$) of the circularly polarized light beam. The
particular feature of that technique is that the transmitted light
intensity is modulated at the frequency $\omega_\mathrm{rf}$ of the
oscillating field, when $\omega_\mathrm{rf}$ is tuned close to
$\omega_L$. The amplitude of the modulation depends as $\sin
2\theta$ with $\cos \theta=\widehat{B_0}\cdot\widehat{k}$. The
amplitude and phase of the modulation depend on $\omega_\mathrm{rf}$
as a classical Lorentz oscillator with a resonance frequency
$\omega_L$. On resonance the phase shift between the oscillating
$B_1$-field and the transmitted light modulation is $90^\circ$ and
for a small detuning $\delta\omega=\omega_\mathrm{rf}-\omega_L$ the
phase shift varies linearly with $\delta\omega$. The width of the
resonance(s) are determined by the transverse relaxation rate of the
spin polarization, which is limited by several effects. In atom-atom
collisions only the sum of the angular momenta of the collision
partners is preserved but spin-exchange processes can change the
individual polarizations. The rate of spin exchange depolarization
is proportional to the collision rate, i.e., to the vapor density,
the spin-exchange cross-section, and the relative velocities of the
collision partners. The dominant depolarization mechanism is due to
collisions of the atoms with the cell walls and the depolarization
rate depends on the adsorption time of the atoms on the walls and on
the wall collision rate. This process can be suppressed by either
preventing the atoms from reaching the walls through the addition of
an inert buffer gas or by reducing the sticking time on the walls by
a suitable coating of the wall surfaces by paraffin or silanes.
However, these coatings may act as a sink for alkali atoms thereby
significantly lower the atomic density \cite{Alex_LIAD}. A stable
vapor pressure is established by having the vapor in thermal
equilibrium with a droplet of alkali metal contained in a sidearm.
Depolarizing collisions with the bulk metal are suppressed by
connecting the cell proper to the sidearm via a small aperture.

The intrinsic linewidth resulting from the combined action of the
mentioned depolarization effects depends on temperature, quality of
the wall coating and cell geometry. Besides those intrinsic
broadening mechanisms the interaction with the optical field and
with the oscillating magnetic field further broaden the magnetic
resonance line. These processes are known as optical and r.f. power
broadening respectively.

\subsection{Effects of hyperfine structure}
\label{sec:EffectsofHFS}

The ground state of alkali atoms with a nuclear spin $I$ splits into
two hyperfine levels with total angular momenta $F_\pm=I\pm 1/2$
with $2F_\pm +1$ Zeeman sublevels labeled by the magnetic quantum
number $M$, respectively. The general evolution of the hyperfine
levels in a magnetic field is described by the Breit-Rabi-Formula
\cite{BreitRabi}. In low magnetic fields (Zeeman interaction $\ll$
hyperfine interaction) the energy of the state $|F_\pm,M\rangle$ is
shifted by $\Delta E_{\pm,M}=g_\pm \mu_B B_0 M$. Here, $\mu_B$ is
the Bohr magneton, and the g-factors $g_\pm$ are given by
\begin{eqnarray}
  \nonumber g_+ &=& +\frac{1}{2I+1}g_J-\frac{2I}{2I+1} g_I  \\
  g_- &=& -\frac{1}{2I+1}g_J-\frac{2I+2}{2I+1} g_I\,,
  \label{eq:gplusminus}
\end{eqnarray}
where $g_J>0$ is the electronic g-factor, defined via
$\vec\mu_J=-g_J \mu_B \vec J/\hbar$ and $g_I$ is the nuclear
g-factor, defined via $\vec\mu_I=g_I \mu_B \vec I/\hbar$. The
magnetic resonance process consists in driving transitions between
adjacent sublevels with a resonance frequency $\omega_L$ given by
\begin{equation}\label{eq:nuL}
    \nu_L=\left|\frac{\Delta E_{\pm,M+1}-\Delta E_{\pm,M}}{h}\right|=\left|\frac{g_\pm \mu_B
    B_0}{h}\right|\,,
\end{equation}
which is equivalent to Eq.~\ref{eq:Larmorfrequency} with
\begin{equation}\label{eq:gammaFdefinition}
    \gamma_F=g_\pm \mu_B /\hbar\,.
\end{equation}
For $^{133}$Cs one has
\begin{eqnarray}
  \nonumber \gamma_4/ 2 \pi &=&+3.4986\,\mbox{Hz/nT}  \\
  \gamma_3/2 \pi &=&-3.5098\,\mbox{Hz/nT}\,.
  \label{eq:gplusminusnumerical}
\end{eqnarray}

In second order in the field $B_0$ the levels acquire an
additional energy shift depending on $M^2$ and $B_0^2$ (quadratic
Zeeman effect) which shifts the $|F_\pm,M\rangle \rightarrow
|F_\pm,M+1\rangle$ transition frequency by an additional amount
\begin{equation}\label{eq:QuadrZeeman}
    |\Delta \nu^{(2)}_L|=\left(\frac{(g_J +g_I) \mu_B
    B_0}{h (2I+1)
    }\right)^2\,\frac{2M+1}{\nu_\mathrm{hfs}} \\
    =\frac{\epsilon}{2} B_0^2\,,
\end{equation}
where $\nu_\mathrm{hfs}$ is the ground state hyperfine splitting.
The quadratic Zeeman effect thus splits the magnetic resonance into
a series of equidistant lines separated by
\begin{equation}\label{eq:QuadrZeemanSep}
    \delta \nu^{(2)}=\epsilon\, B_0^2,
\end{equation}
For cesium ($I=7/2$), one has
\begin{eqnarray}\label{eq:epsilon}
    \epsilon=2.6716\,\mbox{nHz/nT}^2\,.
\end{eqnarray}

In the 2\,$\mu$T field used here $\delta \nu^{(2)}=0.011$\,Hz ,
which is much smaller than the resonance linewidth. The low field
approximation (Eq.~\ref{eq:Larmorfrequency}) is therefore valid for
the present work.

\subsection{Practical realization}

\begin{figure}
  \includegraphics[scale=1]{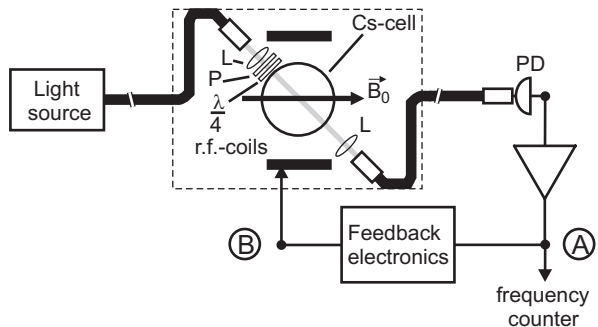}
  \caption{Principle of the $M_x$ magnetometer. The light source
  is the discharge lamp or the diode laser as explained in the
  text. L: lens, P: polarizer and D$_1$ interference filter (in
  case of lamp pumping), $\lambda/4$: quarter-wave plate, PD:
  photodiode.}
  \label{fig:Mag_GenPrinc}
\end{figure}

\begin{figure}
  \includegraphics[scale=1]{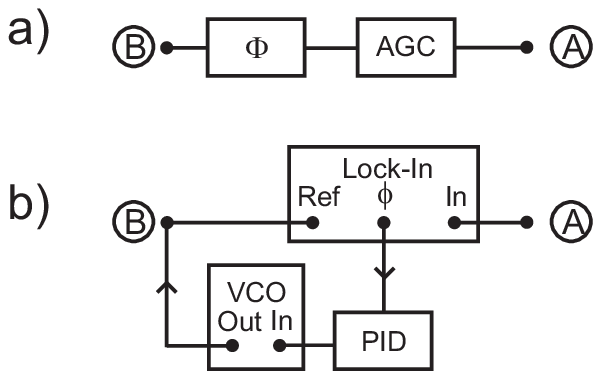}
  \caption{Scheme of feedback electronics in the self-oscillating
  mode (a) and in the phase-stabilized mode (b).
  AGC: amplitude gain control, $\Phi$: phase shifter,
  VCO: voltage-controlled oscillator, PID: feedback controller.}
  \label{fig:PS_SO_Electronics}
\end{figure}
In this work we compare the performance of lamp pumped and laser
pumped magnetometers. Both devices have a common basic design
consisting of the light source, the sensor head, the detector and
feedback electronics (Figs.~\ref{fig:Mag_GenPrinc},
\ref{fig:PS_SO_Electronics}). The sensor head contains the sensor
proper, a spherical glass cell (60\,mm diameter for the LpOPM,
produced in the group of one of the authors, A.S.P., and 70\,mm
diameter for the LsOPM, purchased from MAGTECH Ltd., St.~Petersburg,
Russia) coated with paraffin in which cesium vapor is in thermal
equilibrium with a droplet of metallic cesium at room temperature.
The cell is mounted in a (200\,mm long, 110\,mm diameter)
cylindrical housing. The pumping light is carried from the light
source to the sensor cell by a multimode fiber (800$\,\mu\mbox{m}$
diameter) in the LsOPM and by a fibre bundle (6\,mm diameter) in the
LpOPM. The light transmitted through the cell is carried back to a
detector (photodiode) by an identical fiber in the LsOPM and by an
8\,mm diameter fiber bundle in the LpOPM. The lengths of the fibers
are 8\,m and  5\,m length respectively for the laser and the lamp
pumped device. The sensor head contains also polarization optics
(linear polarizer and quarter-wave plate) for making the light
circularly polarized prior to entering the cell as well as lenses
for collimating the incoming light and focussing the outgoing light
into the return fiber (bundle). Particular care was taken to use
only non-magnetic components in the sensor head. The coils producing
the oscillating field consist of two 70\,mm diameter loops with 12
turns of copper wire each, separated by 52\,mm. When two sensors are
operated in close proximity, the cross-talk of the respective r.f.
fields is avoided by sliding a 1\,mm thick Al cylinder over the
heads. The optical, electronic and mechanical components of the
LsOPM were produced at the University of Fribourg, while the LpOPM
was realized at the Ioffe Institute.

\begin{figure}[ht]
    \includegraphics[scale=1]{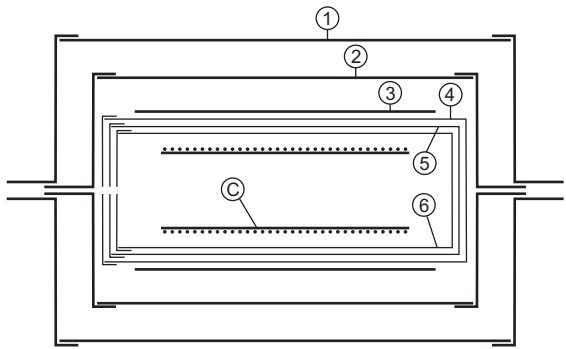}\\
  \caption{Scheme of the Fribourg shield. The labels 1--6 denote
  the different layers as described in
  Tab.~\ref{tab:FribourgShield}. C: magnetic field coil.}
  \label{fig:FribourgShield}
\end{figure}

\begin{table}
  \centering
  \begin{tabular}{|c|c|c|c|c|}\hline
    Layer & $d$ (mm) & $l$ (mm) & $t$ (mm) & Material\\ \hline\hline
    1 & 600 & 900 & 1.5 & Mumetal\\ \hline
    2 & 450 & 750 & 1.5 & Mumetal\\ \hline
    3 & 300 & 600 & 1.5 & Mumetal\\ \hline
    4 & 285 & 743 & 0.76 & Co-Netic\\ \hline
    5 & 256 & 714 & 0.76 & Co-Netic\\ \hline
    6 & 229 & 686 & 0.76 & Co-Netic\\ \hline
  \end{tabular}
  \caption{The Fribourg magnetic shield. $d$: inner diameter,
  $l$: inner length, $t$: layer thickness. Note that all layers
  are closed by endcaps except of layer 3.}
  \label{tab:FribourgShield}
\end{table}

The characterization of the magnetometers described in
Sec.~\ref{sec:IntrSens} was performed in Fribourg. Two magnetometers
were placed inside of a multi-layer cylindrical magnetic shield as
shown in Fig.~\ref{fig:FribourgShield} and
Tab.~\ref{tab:FribourgShield}. The magnetic field of 2\,$\mu$T was
produced by a 50\,cm long, 15\,cm diameter solenoid driven by an
ultra-low noise current supply.

For a direct comparison of the noise performance with an accuracy
below 0.1\,pT the level of magnetic field fluctuations at the
experimental site have to be kept below that level, a performance,
which is hard to realize. Magnetic field variations (in a 1\,Hz
bandwidth) in unshielded environments are of the order of several
$\mbox{nT}_\mathrm{rms}$ or more. A shielding factor exceeding
10'000 is thus needed to suppress fluctuations at that level.

\subsection{Features of the LsOPM and the LpOPM}

\begin{figure}
  \includegraphics[scale=1]{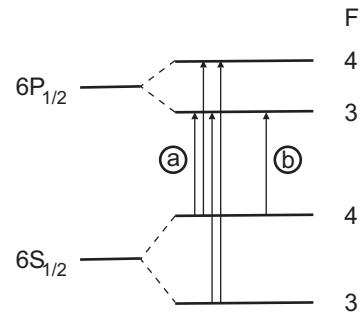}
  \caption{Hyperfine structure of the cesium D$_1$ line.
  The arrows indicate the transitions driven by the discharge
  lamp (a) and the laser (b).}
  \label{fig:CsD1}
\end{figure}

In both types of magnetometers  the D$_1$ transition
$|6S_{1/2}\rangle\rightarrow |6P_{1/2} \rangle$ of Cs at a
wavelength of 894\,nm is used for optical pumping. The LpOPM is
driven by an electrodeless discharge lamp, in which a power
stabilized high frequency generator ($\sim$ 100\,MHz) produces a
discharge in a 12\,mm diameter glass bulb containing cesium vapor
and Xenon as buffer gas. The pumping light is collimated and
filtered by a D$_1$ interference filter centered at 894.5\,nm with a
FWHM=11.5\,nm. Because of the high temperature of the discharge
plasma the spectrum of the emitted D$_1$ radiation is considerably
broader than the Doppler width of the room temperature absorption
line in the sensor cell. All four hyperfine components of the D$_1$
line are excited simultaneously as indicated in Fig.
\ref{fig:CsD1}(a). As the same light is used for detecting the
ground state spin precession the LpOPM detects magnetic resonance in
both the F=4 and the F=3 hyperfine ground states. Because of the
differing g-factors of the two states
(Eq.~\ref{eq:gplusminusnumerical}) the corresponding magnetic
resonance lines are split by 22\,Hz in the 2\,$\mu T$ field, which
is larger than the width of the magnetic resonance lines
($\sim$2.5--5\,Hz) under optimized conditions. The F=3 component is
much weaker than the F=4 component, so that the former plays a minor
role for magnetometry.

The  LsOPM is pumped by a tunable extended cavity laser in Littman
configuration (Sacher Lasertechnik, model TEC500). The output
power of more than 10\,mW exceeds the power required for
magnetometry by more than three orders of magnitude. Therefore a
single laser can be used to drive dozens of magnetometers in
experiments calling for the simultaneous monitoring of the
magnetic field in different locations. The laser frequency is
actively stabilized to the Doppler broadened
$\mbox{F}=4\rightarrow\mbox{F}=3$ hyperfine transition using a
dichroic atomic vapor laser lock (DAVLL) \cite{DAVLL} in an
auxiliary (vacuum) cesium cell. This lock is very stable and
allows a reliable operation of the magnetometer over measurement
periods of several weeks.

In both magnetometers the states $|6S_{1/2};F=4,M=4\rangle$ are dark
states\footnote{Owing to the resolved hyperfine structure in the
LsOPM the states $|6S_{1/2};F=4,M=3\rangle$ and
$|6S_{1/2};F=3,M\rangle$ are also dark states.} which can not absorb
circularly polarized light. At the same time the repopulation of
absorbing (bright) states by the magnetic resonance r.f. transitions
from these levels is used for the optical detection of the magnetic
resonance. The LsOPM drives the $|6S_{1/2};F=4\rangle\rightarrow
|6P_{1/2};F=3\rangle$ transition, and excited atoms can decay into
any of the sublevels of the $|6S_{1/2};F=3\rangle$ hyperfine state,
which do not interact with the narrow band light. This process,
called hyperfine pumping, degrades the efficiency of the optical
pumping process and reduces the overall spin polarization which can
be achieved with laser radiation. A straightforward way to reduce
that loss is to empty the $|6S_{1/2};F=3,\,M\rangle$ sublevels using
a repumping laser tuned to a transition emanating from the F=3
hyperfine ground state. An experimental study of that process is
underway. In the LpOPM all four hyperfine components are pumped
simultaneously. As a result the loss due to hyperfine pumping is
excluded and a larger spin polarization is obtained in the F=4
state. The drawback of the large spectral width of the beam from the
lamp is that an appreciable part of its spectrum lies outside of the
room temperature absorption spectrum of the sensor cell. The
corresponding photons carry no spectroscopic information, but
produce excess shot noise in the detected photocurrent. In that
respect narrow-band laser light leads to a better detection
efficiency in the LsOPM.

\subsection{Modes of operation}
Both types of magnetometers were operated in two different modes.
The self-oscillating mode (SOM) (Fig.~\ref{fig:PS_SO_Electronics} a)
uses the fact that at resonance the driving r.f. field and the
modulated photocurrent are dephased by 90$^\circ$. For that reason
the sinusoidal part of the photocurrent can be used, with an
appropriate amplification and phase shift to drive the r.f. coils in
a feedback loop. In such a configuration the system will
auto-oscillate at the Larmor frequency. An amplitude gain control
ensures that the amplitude of the coil current is kept constant
independently of phase and frequency, so that any r.f. power
dependent systematic effects are suppressed.

The phase-stabilized mode (PSM) (Fig.~\ref{fig:PS_SO_Electronics} b)
also uses the characteristic phase dependence between the applied
oscillating field and the modulation of the detected photocurrent
for locking the frequency of an external oscillator to the Larmor
frequency. The phase, the in-phase component, and the quadrature
component are detected simultaneously by a lock-in amplifier
(Stanford Research Systems SR830). Both the in-phase signal
(dispersive Lorentzian) and the 90$^\circ$ phase shifted phase
signal (arctan-dependence) show a linear zero-crossing near zero
detuning ($\omega_\mathrm{rf}=\omega_L$). Either of the two signals
can thus be used as discriminant in a feedback loop, which
stabilizes the phase to 90$^\circ$. It can be shown that from a
statistical point of view both signals yield an equivalent magnetic
field sensitivity. The phase signal is less sensitive to light power
fluctuations, which may be advantageous to suppress systematic
effects related to power fluctuations \cite{georgJOSA}. However, the
used commercial digital lock-in amplifier had only a moderate update
rate of 400\,Hz of its phase output, so that the much faster
in-phase signal was used in the feedback loop.

In principle the PSM can be understood as a variant of the SOM, in
which the phase-detector, the VCO, and the feedback controller form
a tracking filter. In both modes of operation changes of the
magnetic field lead to instantaneous changes of the Larmor frequency
and thus to instantaneous changes of the transmitted modulation
frequency. The time needed for the radio-frequency to adjust to a
new value after a sudden field change depends on the filters and
other delays in the feedback loop. If a very fast response is not
required, as in our applications, the bandwidth can be decreased by
appropriate filters. In the SOM it is the preamplifier of the
photodetector which limits the bandwidth to 10\,kHz, whereas in the
PSM the feedback loop filters provide a bandwidth up to 1\,kHz
\cite{FRAPLsOPM}. From a practical point of view the registered
bandwidth is limited by the data acquisition system. Because of the
frequency dependence of the phase-shifter the SOM is, in general,
optimized only for a given Larmor frequency, which reduces the
dynamic range of the SOM device. The phase-stabilized magnetometer
keeps the phase-shift at $90^\circ$ independently of the Larmor
frequency. In practice the dynamic range is limited by the frequency
range of the voltage-controlled oscillator (VCO) used to generate
the oscillating field. The long-term stability of both feedback
schemes is limited by temperature dependent phase drifts.

\begin{figure}
  \includegraphics[scale=1]{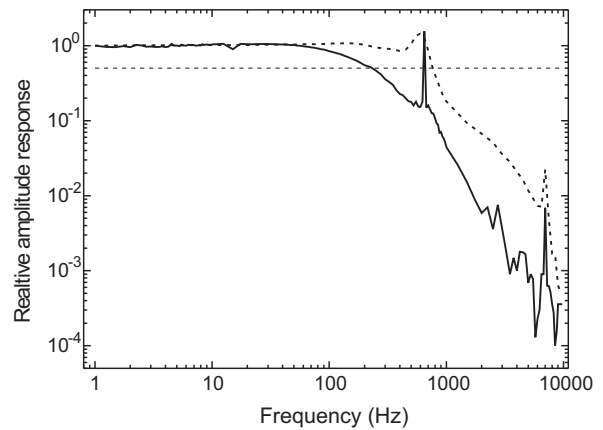}\\
  \caption{Normalized amplitude response of the LsOPM (PSM)
  to a periodic magnetic field change, measured at the VCO
  input: with Al shield (full line) and without Al shield (dashed
  line). The horizontal dashed line indicates an amplitude response
  of 0.5. In both cases the loop filter was adjusted to be about
  750\,Hz. By using the Al shield the bandwidth is reduced to about
  250\,Hz. Note that using the phase-stabilized LpOPM yields the
  same result since the bandwidth is independent of the light source.}
  \label{fig:BWdeponAlshield}
\end{figure}

The Al shield slipped over each magnetometer for avoiding cross-talk
effectively acts as a low-pass filter for external magnetic field
fluctuations (skin effect) which reduces the response bandwidth of
all devices to about 250\,Hz (Fig.~\ref{fig:BWdeponAlshield}).

\section{Performance}

\subsection{Magnetometric sensitivity: Basics and fundamental limit}
\label{sec:IntrSens}

The sensitivity of the magnetometer is defined as the noise
equivalent magnetic flux density (NEM), which is the flux density
change $\delta\!B$ equivalent to the total noise of the detector
signal. In a perfectly stable external magnetic field the smallest
detectable field changes are limited by the intrinsic magnetometer
noise $\delta\!B_\mathrm{int}$. For a measurement bandwidth
$\Delta\nu_\mathrm{bw}$ the intrinsic resolution
$\delta\!B_\mathrm{int}$ depends on the magnetic resonance
linewidth $\Delta\nu$ (HWHM) and on the signal-to-noise ratio
$S/N_\mathrm{int}$ of the magnetometer signal according to
\begin{eqnarray}
  \delta\!B_\mathrm{int} =
  \frac{1}{\gamma}\times\frac{\Delta\nu}{S/N_\mathrm{int}}\,.
  \label{eq:NEMint}
\end{eqnarray}

For a feedback operated magnetometer using an optically thin medium
($\kappa L \ll 1$) the light power (expressed in terms of
photocurrent) detected after the sensor cell is given by
\begin{equation}\label{eq:transmittedcurrent}
I_\mathrm{pc}=I_\mathrm{in} \exp(-\kappa L)\approx I_\mathrm{in}
(1-\kappa L),
\end{equation}
where $I_\mathrm{in}$ is the incident power, $\kappa$ the resonant
optical absorption coefficient and $L$ the sample length. The
absorption coefficient can be written as
\begin{equation}\label{eq:kappa}
    \kappa=\kappa_0(1+\eta \cos\omega_L t),
\end{equation}
where $\kappa_0$ is the mean absorption coefficient and $\eta$ is
the modulation depth which depends - among others - on the degree of
spin polarization and the amplitude of the oscillating field. We can
then write Eq.~\ref{eq:transmittedcurrent} as
\begin{equation}\label{eq:Ipc}
    I_\mathrm{pc}=I_0+I_m \cos\omega_L t=I_0(1+\xi \cos\omega_L t),
\end{equation}
where $I_0=I_\mathrm{in} (1-\kappa_0 L)$ and $I_m=\eta\kappa_0 L
I_\mathrm{in}=\xi I_0$. The contrast $\xi$ is the ratio of the
modulation amplitude and the average photocurrent in the
approximation $\kappa_0 L\ll 1$. The signal $S$ is given by the rms
value of the oscillating part of the magnetometer signal
\begin{eqnarray}\label{eq:SvsIpc}
S=\xi I_0/\sqrt{2}\,.
\end{eqnarray}
The fundamental limit of the magnetometric sensitivity is obtained
for a shot-noise limited signal, with a noise level
$N_\mathrm{int}=N_\mathrm{SN}=\sqrt{2 e I_0 \Delta \nu_\mathrm{bw}}$
of $I_{\mathrm{pc}}$. The shot-noise limited sensitivity then reads
\begin{eqnarray}
  \delta\!B_\mathrm{SN} =
  \frac{\Delta\nu}{\gamma}\,\frac{2}{\xi}\sqrt{\frac{e\,\Delta\nu_\mathrm{bw}}{I_0}}\,.
  \label{eq:NEMintshot}
\end{eqnarray}
The magnetometer signal of interest is contained in the amplitude
and phase of the sinusoidal modulation of the light power after the
sensor cell. The power  spectrum of the spectral power density of an
ideal magnetometer in a perfectly stable magnetic field thus
consists of a delta function centered at the Larmor frequency,
superposed on a flat background of shot-noise fluctuations. In
practice the peak is broadened by the resolution (1\,Hz) of the FFT
analyzer (Stanford Research Systems, model SR760) used for its
recording. The relevant noise contributions which define the S/N
ratio are fluctuations of the photocurrent at the Larmor frequency,
i.e., the value of the background below the Larmor peak. In practice
that ideal spectrum is modified by various imperfections which
degrade the magnetometer performance. In the following we address
contributions from light power fluctuations and magnetic field
fluctuations.

\subsection{Limitations by light power fluctuations}
\label{sec:powerlimit}

\begin{figure}
    \includegraphics[scale=1]{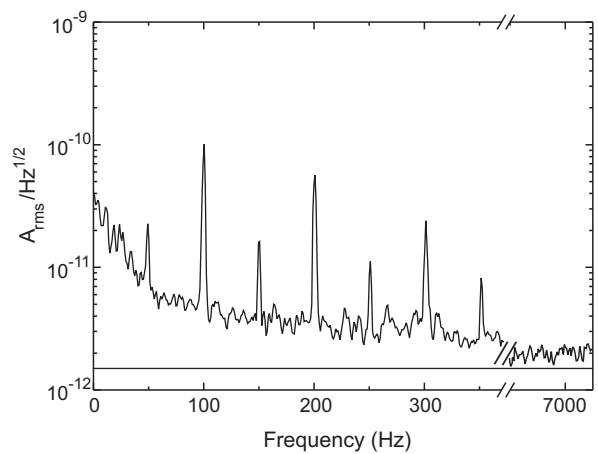}\\
  \caption{Power spectral density of the low frequency noise of
  the laser power. The power was 13\,$\mu$W and the corresponding
  shot-noise level is shown as solid line. The noise at 7\,kHz,
  which lies 50\% above the shot noise level, is also shown.}

  \label{fig:lowfreqlasernoise}
\end{figure}

The light power $I_\mathrm{in}$ has a continuous (technical) noise
spectrum, which lies above the shot-noise level, in particular at
frequencies below 100\,Hz, as shown for the laser in
Fig.~\ref{fig:lowfreqlasernoise}. The individual peaks are even and
odd harmonics of the 50 Hz line frequency. Power fluctuations
contribute to the photocurrent noise at $\omega_L$ by two distinct
processes. First, there is a direct contribution via the noise
component of $I_0$ in Eq.~\ref{eq:Ipc} at the Fourier frequency
$\omega_L$. At $\omega_L/2\pi$= 7\,kHz this noise level is close to
the shot-noise level. The second contribution is due to the second
term in (\ref{eq:Ipc}). Each Fourier component (at $\omega$) of the
power fluctuations is multiplied by $\cos \omega_L t$, and this
mixing produces sidebands at $\omega_L\pm \omega$ in the power
density spectrum. In this way the continuous low frequency part of
the technical noise around $\omega = 0$
(Fig.~\ref{fig:lowfreqlasernoise}) produces a symmetric background
under the Larmor peak. Although the power noise around $\omega=0$ is
18 times (Fig.~\ref{fig:lowfreqlasernoise}) larger than the shot
noise around $\omega=\omega_L$, it is suppressed - according to
Eq.~\ref{eq:Ipc} - by a factor $\xi$, which has a value of
approximately 0.05 in the LsOPM. As a consequence the contribution
of the modulation term in Eq.~\ref{eq:Ipc} to the photocurrent noise
is less than the contribution from $I_0$. We have verified that this
is indeed fulfilled in a carefully calibrated auxiliary experiment.

\subsection{Limitations by magnetic field fluctuations}
\label{sec:fieldlimit}
 In presence of uncorrelated magnetic field
fluctuations, $\delta\!B_\mathrm{ext}$, the highest resolution
with which a magnetic field change can be detected is given by
$\delta\!B=\sqrt{\delta\!B_\mathrm{int}^2+\delta\!B_\mathrm{ext}^2}$.
The fluctuations $\delta\!B_\mathrm{ext}$ of the external magnetic
field can be parameterized by the equivalent noise
$N_\mathrm{ext}$ that they produce on the signal, and $\delta\!B$
can be expressed in a form similar to Eq.~\ref{eq:NEMint} by
\begin{eqnarray}
  \delta\!B =\frac{1}{\gamma}\times\frac{\Delta\nu}{S/N}\,,
  \label{eq:NEM}
\end{eqnarray}
where $N^2=N_\mathrm{int}^2+N_\mathrm{ext}^2$. Fourier components of
the field fluctuations at frequency $\omega$ will mix with the
magnetometer oscillation frequency $\omega_L$ in Eq.~\ref{eq:Ipc}.
Monochromatic field fluctuations, such as the 50\,Hz line frequency
and harmonics thereof produce symmetric sidebands, while low
frequency magnetic field fluctuations produce a continuous
background underlying the Larmor peak.

\subsection{Measurement of the intrinsic sensitivity}
\label{sec:intrLW}

The intrinsic linewidth of the magnetic resonance transition was
measured by extrapolating the experimental linewidth to zero light
power and zero r.f. power. We found a HWHM of 1.63\,Hz for the cell
in the LpOPM and 2.35\,Hz for the cell in the LsOPM. As all cells
were manufactured by the same person the difference of the intrinsic
linewidths is probably due to the slightly larger aperture between
the cell and the sidearm in the case of the LsOPM. After optimizing
the signal-to-noise ratio with respect to light and r.f. power the
(power-broadened) magnetic resonance linewidth (HWHM) is $\Delta
\nu$ = 2.5\,Hz for the LpOPM and 5.0\,Hz for the LsOPM.
\begin{figure}
  \includegraphics[scale=1]{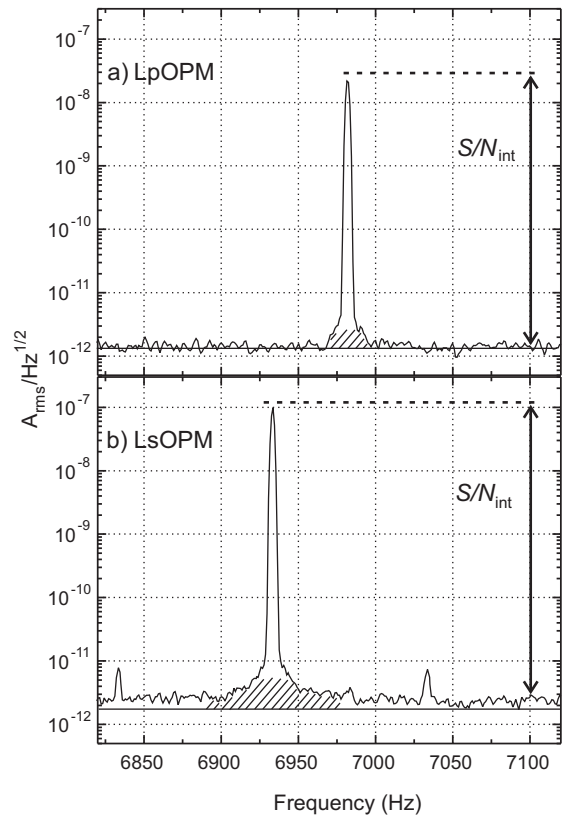}\\
  \caption{Power spectral density plots of the photocurrent
  fluctuations in the LpOPM (a) and the LsOPM (b), both
  operated in the {\em phase-stabilized mode}.
  The signal-to-noise ratios $S/N_\mathrm{int}$ are 29000 (a)
  and 98000 (b) respectively. The photocurrents are 4.3\,$\mu$A
  for the LpOPM and 5\,$\mu$A for LsOPM. The corresponding
  shot-noise levels are represented by the horizontal lines.
  The dashed areas indicate the pedestal discussed in the text.}
  \label{fig:FFTlamplaserPSM}
\end{figure}

\begin{figure}
  \includegraphics[scale=1]{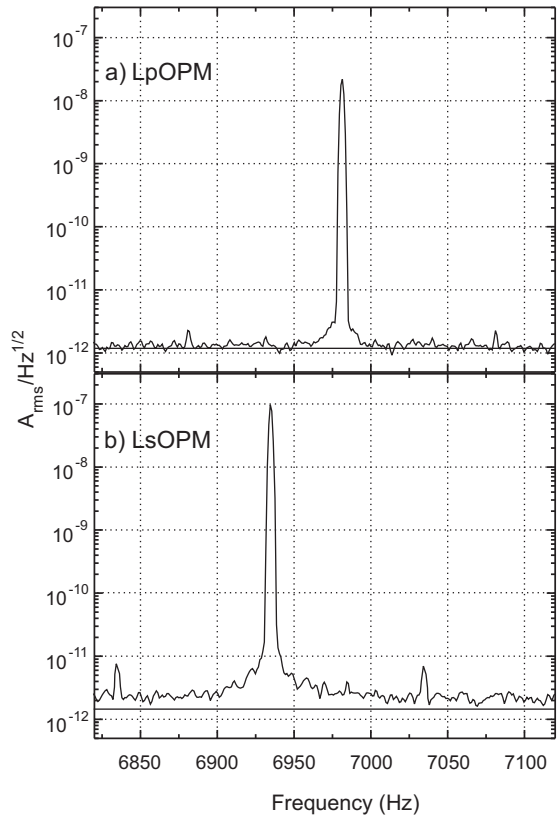}\\
  \caption{Power spectral density plots of the photocurrent
  fluctuations in the LpOPM (a) and the LsOPM (b), both
  operated in the {\em self-oscillating mode}.
  The signal-to-noise ratios are similar to those in
  Fig.~\ref{fig:FFTlamplaserPSM} and the corresponding
  shot-noise levels are represented by horizontal lines.}
  \label{fig:FFTlamplaserSOM}
\end{figure}
Figures~\ref{fig:FFTlamplaserPSM} and \ref{fig:FFTlamplaserSOM} show
typical power density spectra - recorded in the phase-stabilized and
self-oscillating mode of operation - of the r.m.s. voltage
fluctuations at the output of the current-to-voltage preamplifier of
the photodiode current. The spectra contain all structures discussed
above. The pedestal (indicated as dashed area in
Fig.~\ref{fig:FFTlamplaserPSM}) underlying the Larmor peak contains
contributions from fluctuations of the light power and of the
magnetic field. The white noise floor in the far wings of the
central structure represents the intrinsic OPM noise
$N_\mathrm{int}$. Its numerical value for use in Eq.~\ref{eq:NEMint}
was measured 70\,Hz above the carrier. At that frequency this noise
represents the noise component of $I_0$ of the power noise in
Eq.~\ref{eq:Ipc}. We have verified that the noise component of the
modulated contribution in Eq.~\ref{eq:Ipc} is 8 times less than the
noise contribution of $I_0$ under present experimental conditions.
The pedestal under the Larmor peak is thus most probably due to
low-frequency field fluctuations.

After optimizing the magnetometric sensitivity with respect to light
power and r.f. power the LpOPM yields a S/N ratio of $29000$, while
the LsOPM reaches $98000$ in a bandwidth $\Delta\nu_{\mathrm{bw}}$
of 1\,Hz. It can be seen from the figures that the signal-to-noise
ratio does not depend on the mode of operation (SOM or PSM). For the
LpOPM the shot-noise level is nearly reached while in the LsOPM
$N_\mathrm{int}=1.5\times N_\mathrm{SN}$. According to
Eq.~\ref{eq:NEMint} the measured S/N ratios and linewidths under
optimized conditions result a NEM $\delta\!B_\mathrm{int}$ of
$25\,\mbox{fT}$ for the LpOPM and of $15\,\mbox{fT}$ for the LsOPM
in a bandwidth of 1\,Hz. The LsOPM is thus 1.7 times more sensitive
than the LpOPM, although its performance is not yet shot-noise
limited.

\subsection{Discussion}

\begin{figure}
  \includegraphics[scale=1]{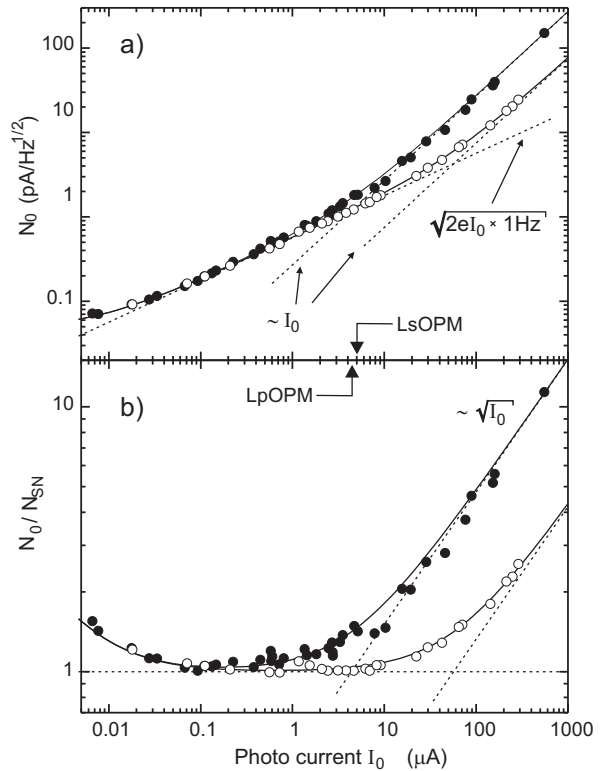}
  \caption{a) Measured noise $N_0$ of the photocurrent produced
  by laser light (black dots) and by light from the discharge
  lamp (open cicles) at 7\,kHz. The measurement bandwidth was 1\,Hz.
  The dashed lines indicate the calculated photodiode shot noise
  ($\propto \sqrt{I_0}$) and the fitted technical laser noise
  ($\propto I_0$). b) Ratio of $N_0$ to the calculated shot noise
  $N_\mathrm{SN}$. The technical noise appears to be proportional
  to $\sqrt{I_0}$ in this representation. The full lines in both
  plots represent the calculated sum of all noise contributions
  according to Eq.~\ref{eq:overalllasernoise}. The arrows refer to
  the (optimized) photocurrent for both devices.}
  \label{fig:lasernoise}
\end{figure}

In order to get a better understanding of the excess power noise in
the LsOPM we measured the dependence of the photocurrent noise on
the light power for the laser in comparison to that of the lamp. In
those measurements only the noise component of $I_0$
(Eq.~\ref{eq:Ipc}) at 7\,kHz was recorded as it represents the
dominant noise contribution of $I_\mathrm{pc}$. For each of the
measurements the cesium cell was removed and the light beam was
detected directly by the photodiode using a transimpedance amplifier
for the photocurrent. The noise $N_0$ measured in this way gives a
lower limit of the intrinsic magnetometer performance. The results
for the laser and lamp source are shown in
Fig.~\ref{fig:lasernoise}. The noise can be written as
\begin{equation}\label{eq:overalllasernoise}
    N_0^2=N_\mathrm{SN}^2+N_\mathrm{dark}^2+N_\mathrm{T}^2,
\end{equation}
where $N_\mathrm{SN}$ is the shot-noise of the photocurrent,
$N_\mathrm{dark}$ the intrinsic detector (photodiode and amplifier)
noise, and $N_\mathrm{T}$ technical noise of the light source, which
is proportional to the photocurrent $I_0$.

$N_\mathrm{dark}$ was measured with the light beams blocked for all
(discrete) amplification stages of the transimpedance amplifier.
With the highest amplification ($10^{-1}\,\mu$A/V) used for
$I_0<600\,\mbox{nA}$ we measured
$N_\mathrm{dark}=48\,\mbox{fA}/\sqrt{\mbox{Hz}}$. This corresponds
to the shot noise of a current $I_0$ of 7\,nA, so that above 7\,nA
the intrinsic detector noise can be neglected. This dark current is
responsible for the deviation of the measured noise from the
(dotted) shot-noise line (Fig.~\ref{fig:lasernoise}) at low
currents.

For larger photocurrents technical noise $N_\mathrm{T}$, which is
proportional to the light power, dominates over the shot noise. We
found $N_\mathrm{T}=k\cdot I_0$, with $k_\mathrm{Ls}=2.6\times
10^{-7}$ and $k_\mathrm{Lp}=0.8\times 10^{-7}$ for the laser and the
lamp source respectively. With the laser source the shot noise
$N_\mathrm{SN}$ becomes equal to the technical noise $N_\mathrm{T}$
for a photocurrent $I_0$ of $4.9\,\mu\mbox{A}$, which thus yields a
noise level $N_0=\sqrt{2} N_\mathrm{SN}$ at that photocurrent. The
laser power for optimized magnetometer parameters corresponds to
$5\,\mu\mbox{A}$, thereby explaining the excess noise of the LsOPM
in Figs.~\ref{fig:FFTlamplaserPSM} and \ref{fig:FFTlamplaserSOM}. In
the case of lamp pumping the technical noise becomes important for
$I_0>60\,\mu\mbox{A}$, so that the magnetometer is shot-noise
limited for the (optimized) photocurrent of 4.3\,$\mu$A.

If one succeeds in eliminating the excess noise of the laser power,
e.g., by an active power stabilization of the LsOPM one can achieve
an intrinsic shot-noise limited sensitivity of $10\,\mbox{fT}$,
thereby outperforming the LpOPM by a factor of $2.5$. This is
compatible with earlier results \cite{AleMx} obtained from a
comparative study of lamp and laser pumped magnetometers using
$^{39}$K, in which the sensitivity of the LsOPM version was found to
be 2.3 times higher than of the corresponding LpOPM device. It is
also interesting to compare those results with the present results
on an absolute scale. The sensitivity of the $^{39}$K-LsOPM was
found to be 1.8\,fT/$\sqrt{\mathrm{Hz}}$. This superior performance
compared to the $^{133}$Cs magnetometer discussed here is mainly due
to the two times larger diameter of the potassium sensor cell,
which, combined with the appreciably smaller spin exchange cross
section of potassium, led to an operating linewidth of 1\,Hz
compared to 5\,Hz with the present Cs-LsOPM. Furthermore the
g-factor of $^{39}$K is twice as large than that of  $^{133}$Cs.
These two factors explain the superior performance of the
K-magnetometer ($1.8\,\mbox{fT}$) compared to the Cs magnetometer
($15\,\mbox{fT}$).

\section{Applications}
\subsection{Direct comparison of LsOPM and LpOPM}
\label{sec:measurements}

\begin{figure}
    \includegraphics[scale=1]{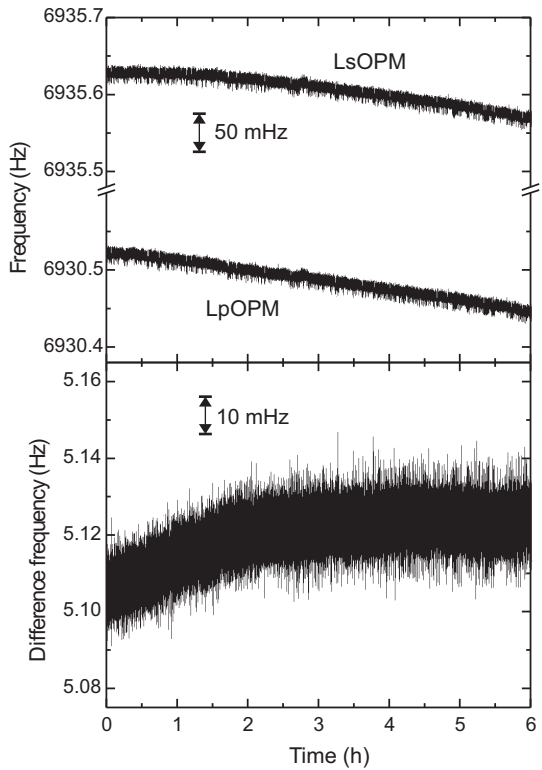}\\
  \caption{Fluctuations of the Larmor frequency recorded with the
  LpOPM (PSM) and the LsOPM (SOM) recorded over 6 hours (top traces).
  The difference of both frequencies shows a drift of the field gradient
  (bottom trace).}
  \label{fig:TSLs2Lp4_6h}
\end{figure}

\begin{figure}
    \includegraphics[scale=1]{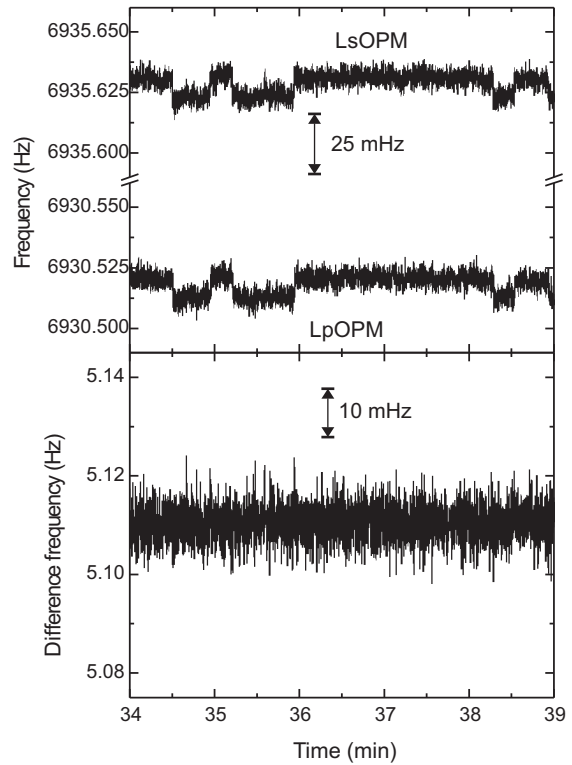}\\
  \caption{Five-minutes slices of the traces in
  Fig.~\ref{fig:TSLs2Lp4_6h}. Individual readings of the
  magnetometers (top) and of their difference frequency
  (bottom).}
  \label{fig:TSLs2Lp4_5min}
\end{figure}

\begin{figure}
    \includegraphics[scale=1]{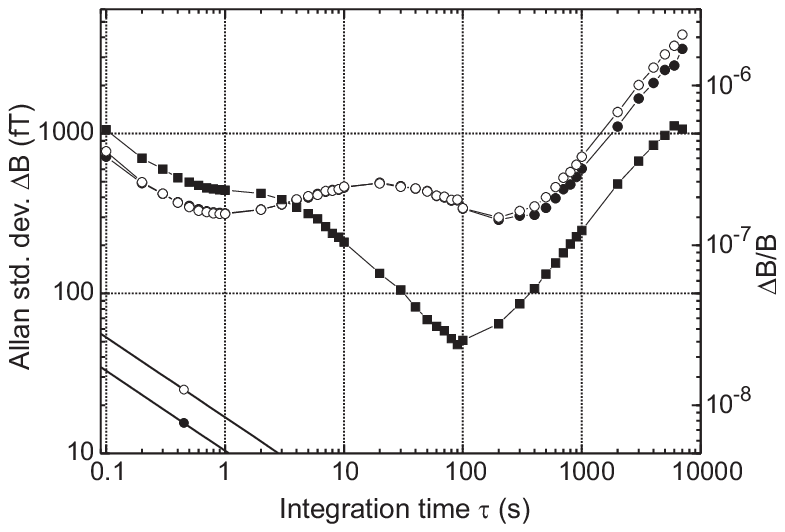}\\
  \caption{Allan standard deviation of the time traces in
  Fig.~\ref{fig:TSLs2Lp4_6h}. Open circles: LpOPM measurement,
  black dots: LsOPM measurement, black squares: field gradient
  measured over 21\,cm. For times below 200 seconds the LsOPM
  data and the LpOPM data points overlap and cannot be distinguished.
  The lines are drawn to guide the eye. In the two lower graphs the
  single points at $\tau=$0.5\,s indicate the NEM of the LpOPM (open
  circle) and of the LsOPM (black dot), and their extrapolation to
  other integration times assuming white noise.}
  \label{fig:ASDLs2Lp4}
\end{figure}

We performed a direct comparison of the performance of the LpOPM and
the LsOPM in simultaneous measurements of magnetic field
fluctuations inside the multilayer shield described above
(Fig.~\ref{fig:FribourgShield}). Both devices were mounted coaxially
in the shield and the centers of their sensor cells were separated
by 21\,cm. The oscillatory signal of each magnetometer was filtered
by a resonance amplifier, centered near 7\,kHz with a FWHM of
500\,Hz, and analyzed by a frequency counter (Stanford Research
Systems, model SR620) with a gate time of 0.1\,s. An example of such
a recording over a continuous interval of 6 hours with the LpOPM
operated in PSM and the LsOPM operated in the SOM is shown in
Fig.~\ref{fig:TSLs2Lp4_6h}. One sees that both devices oscillate a
different average frequencies, which can be explained by the
presence of a magnetic field gradient of 66\,pT/cm which drifts by
0.4\% over the 6 hour interval. The drift is probably due to a
thermal drift of the shield's magnetization. Similar gradients were
measured after interchanging the positions of the two OPMs.
Figure~\ref{fig:TSLs2Lp4_5min} shows a 5 minutes time slice of the
data in Fig.~\ref{fig:TSLs2Lp4_6h}. There are highly correlated
irregular field jumps of approximately 3.6\,pT in both traces. These
fluctuations correspond to relative fluctuations of the solenoid
current at a level of 10$^{-6}$ and are suspected to be caused by
the current source. The Allan standard deviation \cite{barnes} of
the data is a convenient way for characterizing the field drifts on
various time scales. Fig.~\ref{fig:ASDLs2Lp4} shows the Allan plot
of the data from Fig.~\ref{fig:TSLs2Lp4_6h} as a function of
integration time both in absolute and in relative units. Both
magnetometers show the same field stability and the data points are
indistinguishable for small integration times. While the short term
stability is governed by white noise, the bump between 1 and 200\,s
is due to the irregular field jumps. The long term stability is
determined by long term field drifts of the imperfectly shielded
external field and thermal drifts of the solenoid support structure.
At the encountered level of field fluctuations possible false
effects due to the light shift are negligible \cite{FRAPLsOPM}.

\subsection{Active field stabilization}

\begin{figure}
    \includegraphics[scale=1]{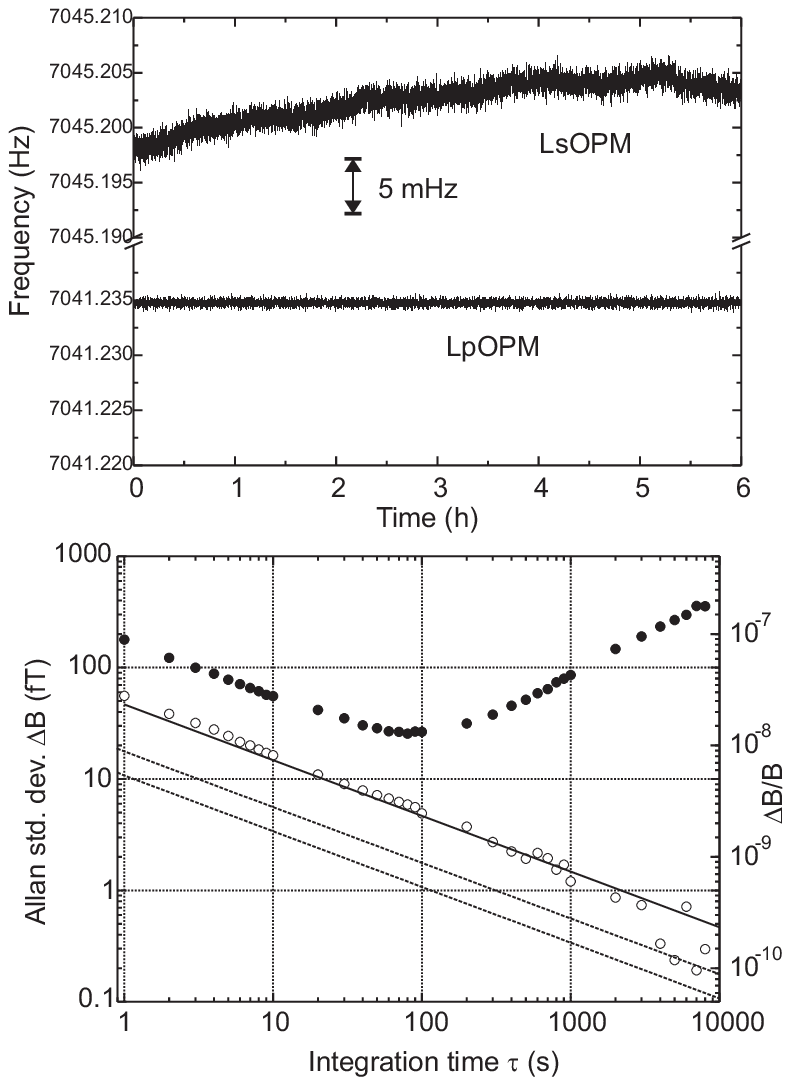}\\
  \caption{Field fluctuations in both sensors when the field is
  stabilized to the LpOPM signal. Upper graph: Time series of the
  residual fluctuations, recorded with a 1\,s gate time of the
  frequency counter. Lower graph: Allan standard deviations of the
  time traces: LpOPM (open circles) and LsOPM (black dots). The solid
  line represents the resolution limit of the frequency counter. The
  dashed lines indicate the NEM of the LpOPM (upper) and of the LsOPM
  (lower).}
  \label{fig:TSASDLp4stab}
\end{figure}

We have further investigated the performance of the magnetometers in
an active magnetic field stabilization system using a phase-locked
loop. For that purpose the phase of the LpOPM oscillation (PSM)
relative to the phase of a reference oscillator was measured by a
lock-in amplifier and used as error signal driving a correction
coil. While field fluctuations common to both sensors are strongly
suppressed by this method, gradient drifts and fluctuations are not
compensated and thus detected by the free-running magnetometer. The
performance of the stabilization scheme is shown in
Fig.~\ref{fig:TSASDLp4stab}, where the top traces represent time
series of the field readings (1\,s gate time of the frequency
counter) of the free-running LsOPM and the LpOPM used for the
feedback. The lower part of Fig.~\ref{fig:TSASDLp4stab} shows the
Allan standard deviation of those data. The jumps in the magnetic
field measured by the LsOPM (Fig.~\ref{fig:TSLs2Lp4_5min}) are
completely suppressed by the feedback loop and only white noise in
the short term stability and contributions from gradient drifts in
the long term stability remain. The field fluctuations reach a
minimum of 25\,fT for $\tau=$100\,s, which corresponds to a relative
field stability of 1.3$\cdot10^{-8}$. Due to the fact that the field
is stabilized to the LpOPM one would expect a significant lowering
of the LpOPM's Allan standard deviation for short integration times.
The white noise behavior (slope $-1/2$ in the Allan plot) of the
LpOPM trace is entirely due to the resolution of the frequency
counter, which is limited by trigger time jitter due to amplitude
noise of the measured sine wave.

We also realized a setup, in which the roles of the LsOPM and the
LpOPM were reversed. The observed performance was identical with the
one described.

\section{Further applications}

OPMs based on Cs vapor are well suited for operation in magnetic
fields smaller than $10\,\mu$T (such as typical fields used in
neutron EDM experiments). In such fields asymmetries of the
magnetic resonance line due to the quadratic Zeeman effect will
not affect their long-term stability (accuracy). In fields between
10\,$\mu$T and 200\,$\mu$T the line asymmetry due to the quadratic
Zeeman effect depends on the light power and power fluctuations
induce fluctuations of that asymmetry and thus of the resonance
frequency. In magnetic fields above 300\,$\mu$T the quadratic
Zeeman shift exceeds the linewidth and the magnetic resonance
spectrum consists of 8 resolved lines \cite{Nett_MultiphotExp}.
This offers a further possibility for the use of cesium
magnetometers. The performance of the LsOPM in that field range is
expected not to be worse than the one measured in a 2\,$\mu$T
field. This expectation is based on the fact that the (performance
determining) laser power noise does not increase with frequency,
so that a shot-noise limited sensitivity can be reached. A
magnetometer based on the quadratic Zeeman effect in potassium has
been demonstrated earlier \cite{Alex_KOPM4q}.

In contrast to buffer gas cells, which require the whole cell volume
to be illuminated in order to achieve a maximum sensitivity,
paraffin-coated cells can be pumped with a laser beam of much
smaller diameter. This allows one to adapt the spatial dimensions of
the paraffin-coated sensors to specific experimental requirements.
One can think, e.g., of using very large cells of several liters for
measuring volume averaged fields. Compared to other proposed large
cell schemes\cite{Heil_He3Mag}, the use of a Cs-OPM offers the
further advantage of a high temporal resolution. Recently, a novel
type of optically pumped magnetometer with a sub-fT (gradiometric)
sensitivity was demonstrated \cite{romalis_Nature}. Besides its use
of very high buffer gas pressures and its operation at a temperature
of 190$^\circ$C, specific features of that magnetometer are its
limited operation range near zero field and its reduced bandwidth of
20\,Hz.

\section{Acknowledgements}

We thank G. Bison for fruitful discussions and careful reading of
this manuscript. We acknowledge financial support from
Schweizerischer Nationalfonds, INTAS, and Paul-Scherrer-Institute
(PSI). One of the authors (A.S.P.) acknowledges financial support
from the University of Fribourg.

\end{document}